\documentclass[12pt]{iopart}
\usepackage{graphicx}

\newcommand{\To}{\longrightarrow}

\begin{document}
\title{A Generic Hopf Algebra for Quantum Statistical Mechanics}%
\author{A. I. Solomon{$^1$ $ ^2$}, G. E. H. Duchamp$^3$, P. Blasiak$^4$, A. Horzela$^4$ and K.A. Penson$^2$}
\address{
$^1$ Physics and Astronomy Department,The Open University,
Milton Keynes MK7~6AA, UK}
\address{
$^2$ Lab.de Phys.Th\'eor. de la Mati\`ere Condens\'ee,
University of Paris VI, France}
\address{
$^3$ Institut Galil´ee, LIPN, CNRS UMR 7030 99 Av. J.-B. Clement, F-93430 Villetaneuse,
France}%
\address{
$^4$ H. Niewodnicza{\'n}ski Institute of Nuclear Physics,
Polish Academy of Sciences,
Division of Theoretical Physics,
ul. Eliasza-Radzikowskiego 152, PL 31-342 Krak{\'o}w, Poland}

\ead{a.i.solomon@open.ac.uk, gduchamp2@free.fr, pawel.blasiak@ifj.edu.pl, andrzej.horzela@ifj.edu.pl, penson@lptl.jussieu.fr}


\begin{abstract}
In this note we present a Hopf  algebra description of a  bosonic quantum model, using the elementary combinatorial elements of Bell and Stirling numbers.  Our objective in doing this is the following.  Recent studies have revealed that  perturbative Quantum Field Theory (pQFT) displays an astonishing interplay between analysis (Riemann Zeta functions), topology
(Knot theory), combinatorial graph theory (Feynman Diagrams) and algebra (Hopf
structure). Since  pQFT is an inherently complicated study, thus  far not exactly solvable and replete with divergences,  the  essential simplicity of the relationships between these areas can
be somewhat obscured. The intention here is to display some of pevious-mentioned structures  in the context of a simple bosonic quantum theory; i.e. a quantum theory
of non-commuting  operators which do not depend on spacetime. The combinatorial properties of these
boson creation and annihilation operators, which is our chosen example, may be described
by graphs, analogous to the Feynman diagrams of pQFT, which we show possess a
Hopf
algebra structure. Our approach is based on
the quantum canonical partition function for a  boson gas.
\end{abstract}
\maketitle
\section{Bell and Stirling numbers}
A basic combinatorial construct is the {\em Bell number}, $B(n)$.
This  gives the number of ways in which $n$ distinct objects may be distributed among  $n$ identical containers, some of which may remain empty. The first few values of $B(n)$ are
$$
B(n)=1,\,1,\,2,\,5,\,15,\,52,\,203,\,\ldots,\quad n=0,\,1,\,2,\,\ldots\,.
$$
  These values grow rapidly, but less so than $n!$

Related to the Bell numbers
are the {\em Stirling numbers} (of the second kind) $S(n,k)$, which
are defined as the number of ways of putting $n$ different objects
into $k$ identical containers, leaving none empty. From the
definition we have
$B(n)=\sum_{k=1}^n S(n,k)$.

A compact way of defining a combinatorial sequence is by means of a {\em generating function}.  For the Bell numbers, the {\em exponential generating function} is given by
\begin{equation}\label{egf}
\sum_{n=0}^\infty {B(n)}\,\frac{x^n}{n!}= \exp(e^{x}-1)
\end{equation}

A slight generalization of the exponential generating function for the Bell numbers is given by that defining the {\em Bell polynomials} $B_n(y)$:
\begin{equation}\label{bpgf1}
 \exp\left(y\bigl(e^{x}-1\bigr)\right)=\sum_{n=0}^\infty B_n(y)\,\frac{x^n}{n!}
\end{equation}
Note that
\begin{equation}\label{bpoly}
   B_n(y)=\sum_{k=0}^n{S(n,k){y^k}}.
\end{equation}
\subsection{Normal order}
Although somewhat unfamiliar to physicists, the Bell and Stirling
numbers are fundamental in quantum theory. This is because they
arise naturally in the {\em normal ordering problem} of bosonic operators.  By the normal ordering operation ${\mathcal N}f(a,a^{\dagger})$ we mean reorder the boson operators in $f(a,a^{\dagger})$ so that all annihilation operators are on the right. For
canonical boson creation and annihilation operators $\{a^{\dagger},a\}$ satisfying  $[a,a^\dag]=1$, the  Stirling numbers of the
second kind $S(n,k)$ intervene through \cite{k1,k2}
\begin{eqnarray}
{\mathcal N}(a^\dag a)^n=\sum_{k=1}^nS(n,k) (a^\dag)^k a^k\,.
\end{eqnarray}
The corresponding Bell numbers $B(n)=\sum_{k=1}^n S(n,k)$ are
simply the expectation values
\begin{equation}\label{Bell}
 B(n)=\langle z|(a^{\dagger}a)^n|z\rangle_{z=1}
\end{equation}
taken in the coherent state defined by
\begin{equation}\label{cs}
 a|z\rangle=z|z\rangle
\end{equation}
for $z=1$. In fact, for physicists, these equations may be taken
as the {\em definitions} of the Stirling and Bell numbers.

\subsection{Graphs}
We now give a graphical representation of the Bell numbers, based on work of Brody, Bender and Meister \cite{ben1,bbm1}, which we have extended in \cite{bbs}.
Consider labelled lines which emanate from a white dot, the
origin, and finish  on a black dot, the vertex. We shall allow
only one line from each white dot but impose no limit on the
number of lines ending on a black dot. Clearly this simulates the
definition of $S(n,k)$ and $B(n)$, with the white dots playing the
role of the distinguishable objects, whence the lines are
labelled, and the black dots that of the indistinguishable
containers. The identification of the graphs for 1, 2 and 3 lines
is given  in Figure 1 below\footnote{The diagrams of our Figure 1 correspond to those of Figure 3 of reference [3], with an interchange of black spots and white spots.   We have regrouped  them to make the relation with B(n) and S(n.k) more transparent.}.
\begin{figure}[h]
\hspace{1cm}
\resizebox{10cm}{!}{\includegraphics{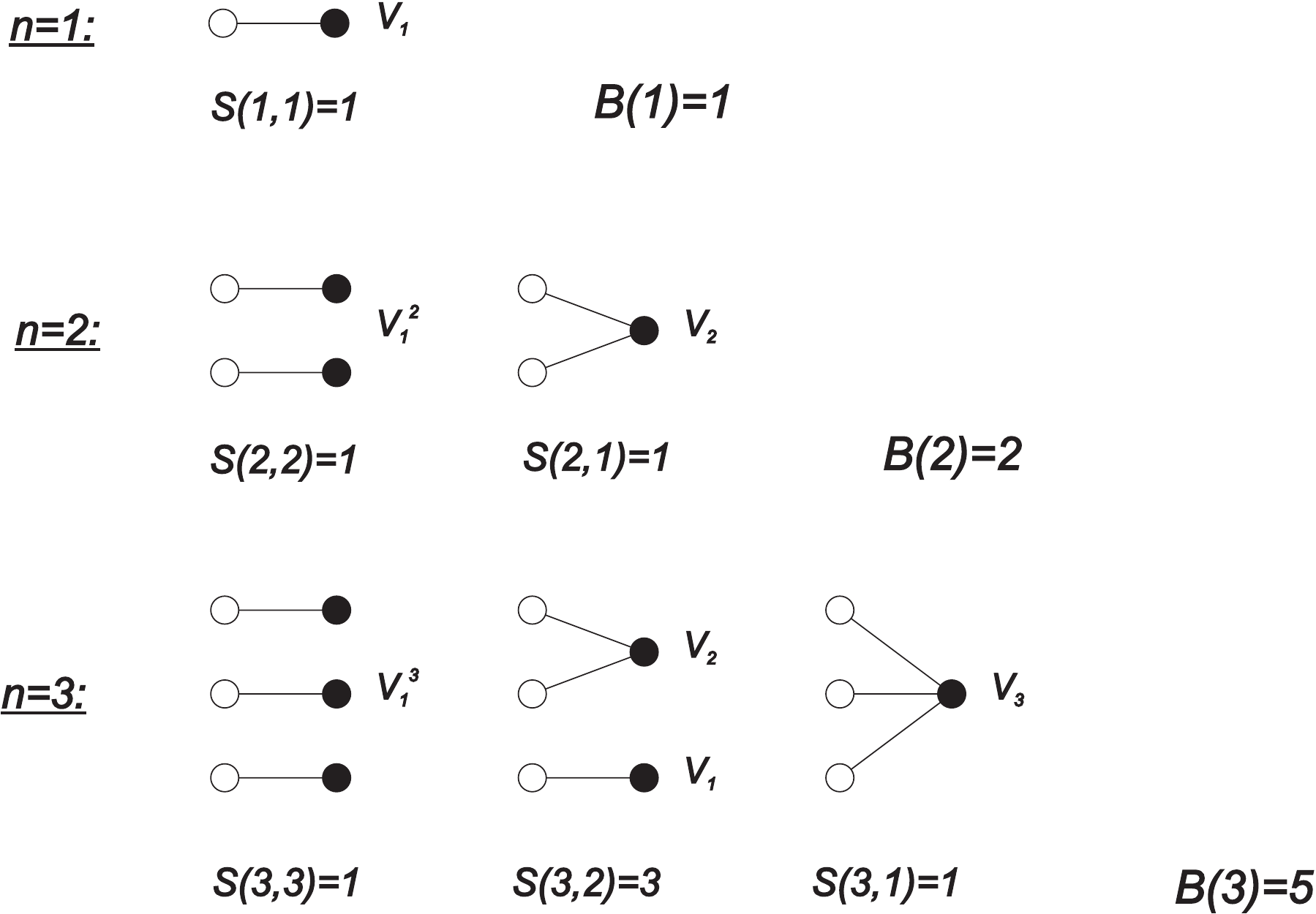}}
\caption{\label{inter}Graphs for $B(n)$, $n=1$, 2, 3.}
\end{figure}
We have concentrated on the Bell number sequence and its
associated graphs since, as we shall show, there is a sense in
which this sequence of graphs is {\em generic} in the evaluation of the Quantum Partition Function. By this we mean the following: We first show below that the Bell number sequence arises from the Partition Function of a {\em non-interacting} boson model. Then,  when interactions are present these may be incorporated by the use of suitable strengths associated with the vertices of the graphs. That is, we can
represent the  combinatorial sequence of an interacting model by the same sequence of
graphs as in  Figure 1, with suitable vertex multipliers
(denoted by the $V$ terms in the same figure).
\section{Partition Function Integrand}
We now show the relation of the preceding considerations to the computation of the Canonical partition Function $Z$ in Quantum Statistical Mechanics, and introduce the Partition Function Integrand, which is directly related to the combinatorial Bell numbers.
\subsection{Free boson gas and Bell polynomials}\label{sub;ho}
The  canonical partition function associated with the hamiltonian $H$ is given by
\begin{equation}\label{pf}
Z={\rm Tr}\exp(-\beta H)\,.
\end{equation}
We take the elementary  case of  the hamiltonian for the single--mode free
boson gas $H=\epsilon a^{\dagger}a$ (ignoring an additive
constant), $\epsilon>0$. The usual computation of the partition
function, exploiting the completeness property
$\sum_{n=0}^{\infty}|n\rangle\langle n|=I$, is immediate:
\begin{eqnarray}\label{pf1}
Z&=& {\rm Tr}\exp(-\beta \epsilon a^{\dagger}a)\\
 &=& \sum_{n=0}^\infty{\langle n|e^{-\beta \epsilon \hat{n}}|n\rangle}\\
 &=& \sum _{n=0}^\infty{e^{-\beta \epsilon n}}\\
 &=&\left(1-e^{-\beta \epsilon}\right)^{-1}\,.
\end{eqnarray}
However, we may use
{\em any} complete set to perform the trace. We choose coherent
states as defined in Eq.(\ref{cs}) above, which are explicitly given by
\begin{equation}\label{cs2}
|z\rangle= e^{-|z|^2/2}\sum_n ({{z^n}{/n!}) {a^{\dagger}}^n}|0\rangle.
\end{equation}
For these states the completeness or {\em resolution of unity}
property is
\begin{equation}\label{cs1}
 \frac{1}{\pi}\int d^{2}z |z\rangle\langle z|=I\equiv
 \int d\mu(z)|z\rangle\langle z|.
\end{equation}
The appropriate trace calculation is
\begin{eqnarray}\label{tr1}
Z&=&\frac{1}{\pi}\int d^{2}z \langle z|\exp\bigl(-\beta \epsilon a^{\dagger}a\bigr)|z\rangle=\\
&=&\frac{1}{\pi}\int d^{2}z \langle
z|:\exp\bigl(a^{\dagger}a(e^{-\beta\epsilon}-1)\bigr):|z\rangle\,,
\end{eqnarray}
where we have used the following well-known relation
\cite{ks,lou} for the {\em forgetful} normal ordering operator
$:\!f(a, a^{\dagger})\!:$ which means ``normally order the
creation and annihilation operators in $f$ {\em forgetting} the
commutation relation $[a,a^{\dagger}]=1$''\footnote{Of course,
this procedure may alter the value of the operator to which it is
applied.}:
\begin{equation}\label{no1}
{\mathcal N}exp(xa^{\dagger}a)=:\exp\bigl(a^{\dagger}a(e^{x}-1)\bigr):.
\end{equation}
We therefore obtain, integrating over the angle variable $\theta$
and the radial variable $r=|z|$,
\begin{equation}
Z=\frac{1}{\pi}\int_0^{2\pi}d\theta\int_0^\infty r\,d r
\exp\left(r^2\bigl(e^{-\beta \epsilon}-1\bigr)\right)\,,
\label{z1}
\end{equation}
which gives us $Z=\bigl(1-e^{-\beta\epsilon}\bigr)^{-1}$ as
before.

We rewrite the above equation  to show the connection with our
previously--defined combinatorial numbers. Writing $y=r^2$ and
$x=-\beta \epsilon$, Eq.(\ref{z1}) becomes
\begin{equation}\label{z2}
Z=\int\limits_0^\infty d y\exp\bigl(y(e^{x}-1)\bigr)\,.
\end{equation}
This is an integral over  the classical exponential generating function for
the Bell polynomials as given in Eq.(\ref{bpgf1}). This leads
to the combinatorial form for the partition function
\begin{equation}\label{z3}
Z=\int_0^\infty dy\sum_{n=0}^\infty B_n(y)\,\frac{x^n}{n!}\,.
\end{equation}

Although Eq.(\ref{z3}) is remarkably simple in form, it is often
by no means a straightforward matter to evaluate the analogous
integral for other than the free boson system considered here.
Further, it is also clear that we may not interchange the integral
and the summation, as each individual $y$ integral diverges. We
shall therefore concentrate in  what follows on the {\em partition
function integrand} (PFI) $F(z)=\langle z|\exp(-\beta
H)|z\rangle$, whence $Z=\int F(z)\,d\mu(z)$, to give a graphical
description of a perturbation approach. The function $F$ maps
coherent states $|z\rangle$ to (real) numbers, and is therefore a {\em functional} on the coherent states..
\subsection{General partition functions}
We now apply this graphical approach to the general partition
function in second quantized form. With the usual definition for
the partition function Eq.(\ref{pf}).
In general the hamiltonian is given by  $H=\epsilon
w(a,a^{\dagger})$, where $\epsilon$ is the energy scale, and $w$
is a string (= sum of products of positive powers) of boson
creation and annihilation operators. The partition function
integrand $F$ for which we seek to give a graphical expansion, is
\begin{equation}\label{pfz}
 Z(x)=\int{F(x,z)\,d\mu(z)}\,,
\end{equation}
where
\begin{eqnarray}
F(x,z)&=&\langle z|\exp(xw)|z\rangle= \hskip20mm(x=-\beta \epsilon)\nonumber \\
&=&\sum_{n=0}^{\infty}\langle z|w^n|z\rangle\,\frac{x^n}{n!}\nonumber \\
&=&\sum_{n=0}^{\infty}W_n(z)\,\frac{x^n}{n!}\nonumber \\
&=&\exp\biggl(\;\sum_{n=1}^{\infty}V_n(z)\,\frac{x^n}{n!}\biggr),
\end{eqnarray}
with obvious definitions of $W_n$ and $V_n$. The sequences
$\left\{W_n\right\}$ and $\left\{V_n\right\}$ may each be
recursively obtained from the other \cite{pou}. This relates the
sequence of multipliers $\{V_n\}$ of Figure 1 to the hamiltonian
of Eq.(\ref{pf}). The lower limit $1$ in the $V_n$ summation is a
consequence of the normalization of the coherent state
$|z\rangle$.

At this point the analogue of this model to perturbative quantum field theory becomes more transparent.  For example, to calculate the partition Function $Z(x)$, we must integrate over the Partition Function Integrand $F(x,z)$.  Here, the integration over the coherent state parameter $z$ plays the role of the space-time integration of Quantum Field Theory. As noted above with reference to Eq.(\ref{z3}) for the non-interacting case, and as in pQFT, term-by-term integration, corresponding to fallaciously interchanging the summation and integration actions, results in infinities at each term, as can be directly verified even in the non-interacting case.  Thus even here an elementary form of renormalization is necessary, if we have to resort to term-wise integration.

Considerations such as these have led many authors, as noted in reference \cite{kre}, to consider a global, algebraic approach to pQFT; and this we  now do in the context of our simple model.

\section{Hopf Algebra structure}
To describe the Hopf Algebra structure of our model, which we shall refer to as BELL below, we first introduce a basic Hopf Algebra, which we call POLY, generated by a single parameter $x$.  This is essentially a standard one-variable polynomial algebra on which we impose the Hopf operation of co-product, co-unit and antipode, as a useful pedagogical device for describing their properties. The Hopf structure associated with our model described above is simply a multi-variable extension of {\bf POLY}.
\subsection{The Hopf Algebra {\bf POLY}}
{\bf POLY} consists of polynomials in $x$ (say, over the real  field  $\mathcal R$, for example).The standard algebra structure of addition and associative multiplication is obtained in the usual way, by polynomial addition and multiplication.  The additional Hopf operations are:
\begin{enumerate}
\item The coproduct $\Delta:{\bf POLY}\To {\bf POLY}\times {\bf POLY}$ is defined by
\begin{eqnarray}
\Delta(e)&=&e\times e  \; \; \; \;({\rm unit}\; \; e) \nonumber \\
\Delta(x)&=&x \times e +e \times x  \; \; \; \; ({\rm generator}\; \; x) \nonumber \\
\Delta(AB)&=&\Delta(A)\Delta(B) \; \; \; {\rm otherwise} \nonumber
\end{eqnarray}
so that $\Delta$ is an algebra homomorphism.
\item The co-unit $\epsilon$ satisfies $\epsilon(e)=1$ otherwise $\epsilon(A)=0$.
\item The antipode ${\mathcal S}:{\bf POLY}\To {\bf POLY}$ satisfies ${\mathcal S}(e)=e$; on the generator $x$, ${\mathcal S}(x)=-x$. It is an {\em anti-homomorphism}, i.e. ${\mathcal S}(AB)={\mathcal S}(B){\mathcal S}(A)$.
\end{enumerate}
It may be shown that the foregoing structure ${\bf POLY}$ satisfies the axioms of a commutative, co-commutative Hopf algebra.
\subsection{{\bf BELL}}
We now briefly describe the  Hopf algebra   ${\bf BELL}$ which is appropriate for the diagram structure introduced in this note, defined by  the diagrams of Figure 1.
\begin{enumerate}
\item Each distinct diagram is an individual basis element of ${\bf BELL}$; thus the dimension is infinite. (Visualise each diagram in a ``box''.) The sum of two diagrams is simply  the two boxes containing the diagrams. Scalar multiples are formal; for example, they may be  provided by the $V$ coefficients.
\item The identity element $e$ is the empty diagram (an empty box).
\item Multiplication is the juxtaposition of two diagrams within the same ``box''. ${\bf BELL}$ is generated by the {\em connected} diagrams; this is a consequence of the Connected Graph Theorem \cite{FU}.  Since we have not here specified an order for the juxtaposition, multiplication is commutative.
\item The coproduct $\Delta:{\bf BELL}\To {\bf BELL}\times {\bf BELL}$ is defined by
\begin{eqnarray}
\Delta(e)&=&e\times e  \; \; \; \;({\rm unit}\; \; e) \nonumber \\
\Delta(y)&=&y \times e +e \times y  \; \; \; \; ({\rm generator}\; \; y) \nonumber \\
\Delta(AB)&=&\Delta(A)\Delta(B) \; \; \; {\rm otherwise} \nonumber
\end{eqnarray}
so that $\Delta$ is an algebra homomorphism.
\item The co-unit $\epsilon$ satisfies $\epsilon(e)=1$ otherwise $\epsilon(A)=0$.
\item The antipode ${\mathcal S}:{\bf BELL}\To {\bf BELL}$ satisfies ${\mathcal S}(e)=e$; on a generator $x$, ${\mathcal S}(y)=-y$. It is an {\em anti-homomorphism}, i.e. ${\mathcal S}(AB)={\mathcal S}(B){\mathcal S}(A)$.
\end{enumerate}
\subsection{{\bf BELL} as an extension of {\bf POLY}}
It may be seen that ${\bf BELL}$ is a multivariable version of ${\bf POLY}$. To show this, we  code the diagrams by letters. We use an infinite alphabet $Y=\{y_1,y_2,y_3\cdots \}=\{y_k\}_{k\geq 1}$ and code each connected diagram with one black spot and $k$ white spots with the letter $y_k$. An unconnected diagram will be coded by the product of its letters.\\
In this way, the diagrams of Figure 1 are coded as follows:
\begin{itemize}
	\item first line : $y_1$
	\item second line : $y_1^2$ and $y_2$
	\item third line : $y_1^3$ and $y_1y_2$ and $y_3$\ .
\end{itemize}
Thus one sees that each diagram of weight $n$ with $k$ connected components is coded bijectively by a monomial
of weight $n$ (the weight of a monomial $y_{i_1}y_{i_2}\cdots y_{i_r}$ is just the sum of the indices $\sum_{j=1}^r i_j$) and $k$ letters. The algebra ${\bf BELL}$ is coded by commutative polynomials in the infinite alphabet $Y$; that is, the coding is an isomorphism ${\bf BELL}\rightarrow {\mathcal R}[Y]$. As an aside, one may note that  the basis elements of ${\bf BELL}$ are sometimes referred to as {\em forests}.
As above, it  may be shown that the foregoing structure ${\bf BELL}$ satisfies the axioms of a commutative, co-commutative Hopf algebra.
\section{Discussion}
The objective of this note was to introduce a very simple Hopf structure associated with standard bosonic creation and annihilation operators, in the context of the evaluation of the canonical partition function of quantum statistical mechanics. We did this via a diagrammatic description applied to a non-interacting boson gas, but implied that the algebraic description was general enough to include interactions as scalar coefficients. The inspiration for this task arose from considering recent work on perturbative quantum field theory (pQFT)\cite{kre}, where it has been  shown that a Hopf description is available for this far more complicated system. It is instructive to consider a much more straightforward system, such as the one treated  here, where the operators do not depend on space or time. Nevertheless, we have shown that even such a   basic system does exhibit some of the features of the more complicated case, in particular the structure of a  Hopf algebra, which we called
${\bf BELL}$. This may be thought of as a simple solvable model in its own right.

However, one may also ask  wherein does this simple structure sit within the full pQFT structure?  The strategy which we adopt, and which we describe in further work,  is to generalize the {\em algebraic} structure, and thereby produce Hopf algebras of sufficient complexity to emulate those associated with pQFT. In subsequent work we show that by suitable deformation of the Hopf algebra described here we obtain structures related to those arising from more realistic models, such as those associated with pQFT.
\section*{Acknowlegements} The authors wish to acknowledge support from the Agence Nationale de la Recherche (Paris, France) under Program No. ANR-08-BLAN-0243-2 and from PAN/CNRS Project PICS No.4339(2008-2010) as well as the Polish Ministry of Science and Higher Education Grant No.202 10732/2832.

\end{document}